\documentclass[galaxies, article, accept, moreauthors, pdftex]{mdpi} 

\firstpage{1} 
\pubvolume{1}
\issuenum{1}
\articlenumber{5}
\pubyear{2020}
\copyrightyear{2020}
\history{Received: 27 February 2020; Accepted: 29 March 2020}

\pdfoutput=1

\Title{Integral Field Spectroscopy of Planetary Nebulae with MUSE}

\Author{Jeremy R. Walsh$^{1}$ \orcidA{} and Ana Monreal-Ibero$^{2,3}$ \orcidB{}
}

\AuthorNames{Jeremy Walsh and Ana Monreal-Ibero}

\address{
$^{1}$ \quad Jeremy R. Walsh, European Southern Observatory, 85748 Garching, Germany; jwalshl@eso.org\\
$^{2}$ \quad Instituto de Astrof\'{i}sica de Canarias (IAC), E-38205 La Laguna, Tenerife, Spain; amonreal@iac.es\\
$^{3}$ \quad Universidad de La Laguna, Dpto. Astrof\'{i}sica, E-38206 La Laguna, Tenerife, Spain
}

\corres{Correspondence: jwalsh@eso.org}

\abstract{
The Multi-Unit Spectroscopic Explorer (MUSE) is a large integral field unit
mounted on the ESO Very Large Telescope. Its spatial (60 arcsecond field) 
and wavelength (4800-9300\AA) coverage is well suited to detailed imaging
spectroscopy of extended planetary nebulae, such as in the Galaxy. An overview
of the capabilities of MUSE applied to planetary nebulae (PNe) is provided 
together with the specific advantages and disadvantages. Some examples of 
archival MUSE observations of PNe are provided. MUSE datacubes for two targets 
(NGC~3132 and NGC~7009) have been analysed in detail and they are used to show 
the advances achievable for planetary nebula studies. Prospects for further MUSE 
observations of PNe and a broader analysis of existing datasets are outlined.
}


\begin{document}

\section{Overview of MUSE}
The Multi-Unit Spectroscopic Explorer (MUSE) is a large field of view ($\sim$60$  \times$
60$''$) optical integral field spectrometer mounted on the European Southern Observatory 
(ESO) Very Large Telescope (VLT), currently on Unit Telescope 4 (Yepun). The field is 
divided into 24 slices and each is sent to a separate integral field unit (IFU) that 
divides the sub-field into 48 mini slits which are all fed to one of the 24 identical 
spectrometers \cite{Bacon2010}. Each spectrometer module is equipped 
with a Volume Phase Holographic Grating covering the full wavelength range 4600--9300\AA\ 
and imaged by an EEV deep depletion, anti-reflection coated 4k$\times$4k charge coupled 
detector (CCD). Details of the instrument can be found on the
\href{http://www.eso.org/sci/facilities/paranal/instruments/muse/inst.html}{ESO MUSE webpage}
and in the MUSE Instrument Manual \cite{Richard2019}.

Although the instrument format is fixed in terms of number of slicers and
spectrographs, there are some options for feeding the field from the VLT 
to the instrument and one wavelength range choice. Table \ref{tab:MUSE_inst}
lists the various options. In Wide Field Mode (WFM), the full field of 
300$\times$300 pixels, each 0.20$\times$0.2$''$, is covered. Using the VLT 
Deformable Secondary Mirror (DSM), a Ground Layer Adaptive Optics (AO) 
feed is available which provides the same field but an improved image
quality (typically $\times$2 ensquared energy, but depending on 
atmospheric conditions). Higher spatial resolution is achievable with
Laser Tomographic AO and the field size is reduced to 7.5$''$ (Narrow
Field Mode, NFM).
It should be noted that, since a sodium laser is employed for the AO modes,
the spectral region around the Na I doublet (5890,5896\AA) is blocked by 
a filter. In both WFM and NFM, an alternative wavelength range can be assessed
(called extended mode) which shifts the blue cut-off from 4800 to 4650\AA;
however this mode suffers 2nd order contamination at wavelengths above 8000\AA.    
Most of these modes have been employed for MUSE observations of PNe.

\begin{table}
\caption{MUSE Instrumental Parameters}
\centering
\begin{tabular}{llll}
\hline\hline
Mode     & Field ($''$) & Pixel ($''$) & Image quality \\
\hline
Wide Field Mode (WFM) & 59.9$\times$x60.0 & 0.20 & Native seeing \\
Wide Field Mode + AO  & 59.9$\times$x60.0 & 0.20 & $^{>}_{\sim}$ 2 $\times$ encircled energy \\ 
Narrow Field Mode (NFM) & 7.42$\times$7.43 & 0.025 & 55 -- 80 milli-arcseconds \\
\hline
Spectral mode & Wavelength range (\AA) & Resolving Power & Field mode \\
\hline
Normal (N)    & 4800 -- 9300 & 1770 -- 3590 & WFM \\
Normal (N)    & 4800 -- 5780, 6050 -- 9300 & 1740 -- 3450 & NFM \\
Extended (E)  & 4650 -- 9300 & 1770 -- 3590 & WFM \\
Extended (E)  & 4650 -- 5760, 6010 -- 9300 & 1740 -- 3450 & NFM \\
\hline
\end{tabular}
\label{tab:MUSE_inst}
\end{table}

\section{Suitability of MUSE for PNe}
Planetary nebulae are extended emission and continuum nebulae ionized by
a central hot star, mostly representing the shell of gas ejected during the 
previous asymptotic giant branch (AGB) phase of the evolution of a 0.8 -- 8.0
M$_{\odot}$ star. PNe in the Galaxy present projected sizes from fractions
of an arcsecond for compact and/or distant PNe, to almost half a degree,
making them good targets for IFU observation (e.g., \cite{Dopita2017}).   
MUSE with its large IFU field can be considered well matched to spectral imaging
of a range of Galactic PNe. For nebulae where the emission outskirts of the 
field are contained within the MUSE field, direct sky removal is facilitated; 
nebulae much larger can be effectively sky subtracted using an offset sky 
exposure. Details of compact structures in larger nebulae, or more compact 
PNe, are well accessible to the narrow field mode. The native spaxel size 
of 0.20$''$ is matched to sampling good Paranal seeing ($\sim$0.5$''$) and
smaller scale structures can be imaged in detail; while NFM competes with 
Hubble Space Telescope imaging to the 0.1$''$ level. 

The spectral coverage of MUSE provides a good sampling of the rich optical
emission spectrum consisting of recombination and collisionally excited lines 
which have been observed over many decades and have contributed a substantial 
literature on the lines, their parent atoms and ionization mechanisms and 
their use as diagnostics of physical processes (e.g., \cite{FangLiu2011, 
FangLiu2013, Garcia-Rojas2015}). The applicability 
of diagnostic line ratios for dust extinction, electron density ($N_{\rm e}$) and 
temperature ($T_{\rm e}$), measurement and for abundance determination, refined 
through observations of
PNe, have been applied to ionized regions in galaxies at the highest 
observed redshifts. The MUSE coverage to 9300\AA\ allows the Paschen continuum
jump of ionized H (and ionized He) at $\sim$8210\AA\ \cite{Zhang2004} to be 
measured (although compromised in extended mode by 2nd order contamination),
providing another temperature and density probe distinct from those of the 
emission lines.

The wavelength compass does however bring some disadvantages compared to
full optical (3500--10000\AA) and higher resolution spectroscopy:
\begin{itemize}
\item the spectral resolving power (R) is relatively low ($\sim$1800--3600, see 
Table 1), thus not sampling the emission lines at the optimal value
of two times their intrinsic (thermal and Doppler broadened) width (R$>$10000).
On multiple sampled spectra, some restoration is of course possible to assess
line blends below 3--4\AA\ separation;
\item Important diagnostic lines blueward of 4650\AA\ are not included, 
including the [O~II] doublet (3726,3729\AA) for $N_{\rm e}$ measurement, [Ne~III]
lines and the auroral [O~III]4363\AA\ which, by ratio to the 
[O~III]4959,5007\AA\ lines, provides $T_{\rm e}$  for the higher ionization medium. Alternative less bright diagnostic lines are available 
in the MUSE range, involving less major coolants of the gas, such as
$N_{\rm e}$ from [S~II] 6716/6731\AA, [Cl~III] 5517/5537\AA\ and [Ar~IV] 4711/4740\AA\
in extended mode, $T_{\rm e}$ from [O~I] 6302/5577\AA, [N~II] 6583/5755\AA\ and 
[S~III] 9069/6311\AA\ and [Ar~III] 7135/5192\AA, thus partially compensating for
the absence of the 'classical' O$^{++}$ diagnostic which samples the
bulk ionization component for all but the lowest ionization PNe;
\item the bright and important (at least for higher ionization PNe) He~II 4686\AA\ 
line is missing from the MUSE normal mode (but included in extended mode). This
absence is partially compensated by the He~II 5411\AA\ line, although $\sim$
10 times fainter; 
\item the [O~II]7319,7320\AA\ and 7329,7330\AA\ doublet is much fainter than
the blue [O~II]3726,3729\AA\ lines but can be used for determination of O$^{+}$
ionic fraction;
\item relatively few optical metal recombination lines (ORLs) are covered in
normal mode (mostly N) and while the 4650\AA\ region important for the brightest
O~II recombination lines is just included in extended mode, no Ne ORLs are available.
\end{itemize}

\section{Archival MUSE observations of PNe}
Some 20 PNe have been observed with MUSE to data at various phases of the
instrument commissioning and also in guest observer (open) time. Table \ref{tab:MUSE_arch}
lists the PNe observed with an indication of the mode(s) employed. Data
for only two PNe (NGC~3132 and NGC~7009, see next section) have been 
published to date, reflecting the large analysis task associated with these
rich data sets encompassing hundreds of detectable emission lines per spaxel.

\begin{table}
\caption{MUSE Archival Observations of Planetary Nebulae}
\centering
\begin{tabular}{lllll}
\hline\hline
Target & Designation         & MUSE   & Obs. Type & Notes \\
       &                     & Mode   &           &       \\
\hline
NGC~6369 & PNG 002.4~$+$05.8 & WFM AO, N & Comm.  & \\
M~1-42   & PNG 002.7~$-$04.8 & WFM, E & GO        & ESO Proposal 097.D-0241 \\
Hf~2-2   & PNG 005.1~$-$08.9 & WFM, E & GO        & 097.D-0241 \\
GJJC~1   & PNG 009.8~$-$07.5 & NFM, N & Comm.     & IRAS~18333-2357 \\
Sa~3-151 & PNG 033.2~$-$01.9 & NFM, N & Comm.     & \\
NGC~6778 & PNG 034.5~$-$06.7 & WFM AO, N & Comm.  & \\
NGC~7009 & PNG 037.7~$-$34.5 & WFM, N & SV        & Walsh et al. (2016,2018) \\
NGC~7009 & PNG 037.7~$-$34.5 & WFM, E & GO        & 097.D-0241 \\
NGC~7009 & PNG 037.7~$-$34.5 & NFM, N & Comm.     & \\
Abell~46 & PNG 055.4~$+$16.0 & WFM, E & GO        & 097.D-0241 \\
IC~418   & PNG 215.2~$-$24.2 & WFM, N & Comm.     & \\
NGC~3242 & PNG 261.0~$+$32.0 & WFM, E & GO        & 097.D-0241 \\
NGC~2818 & PNG 261.9~$+$08.5 & WFM, N & Comm.     & \\ 
NGC~3132 & PNG 272.1~$+$12.3 & WFM, N & Comm.     & Monreal-Ibero\&Walsh (2020) \\
IC~2501  & PNG 281.0~$-$05.6 & WFM, N & Comm.     & \\
NGC~4361 & PNG 294.1~$+$43.6 & WFM, N & Comm.     & \\
NGC~5189 & PNG 307.2~$-$03.4 & WFM, N & Comm.     & \\
IC~4406  & PNG 319.6~$+$15.7 & WFM, N & Comm.     & Type I PN, See Fig. 1 \\
Mz-3     & PNG 331.7~$-$01.0 & NFM, N & Comm.     & \\
CPD-56~8032 & PNG 332.9~$-$09.9 & NFM, N & Comm.  & He 3-1333 \\
NGC~6153 & PNG 341.8~$+$05.4 & WFM, E & GO        & 097.D-0241 \\
NGC~6563 & PNG 358.5~$-$07.3 & WFM AO, N & Comm.  & See Fig. 1 \\
\hline
\end{tabular}
\label{tab:MUSE_arch}
\end{table}

\begin{figure}
\centering
\resizebox{\hsize}{!}{
\includegraphics[height=5.0truecm,angle=0,clip]{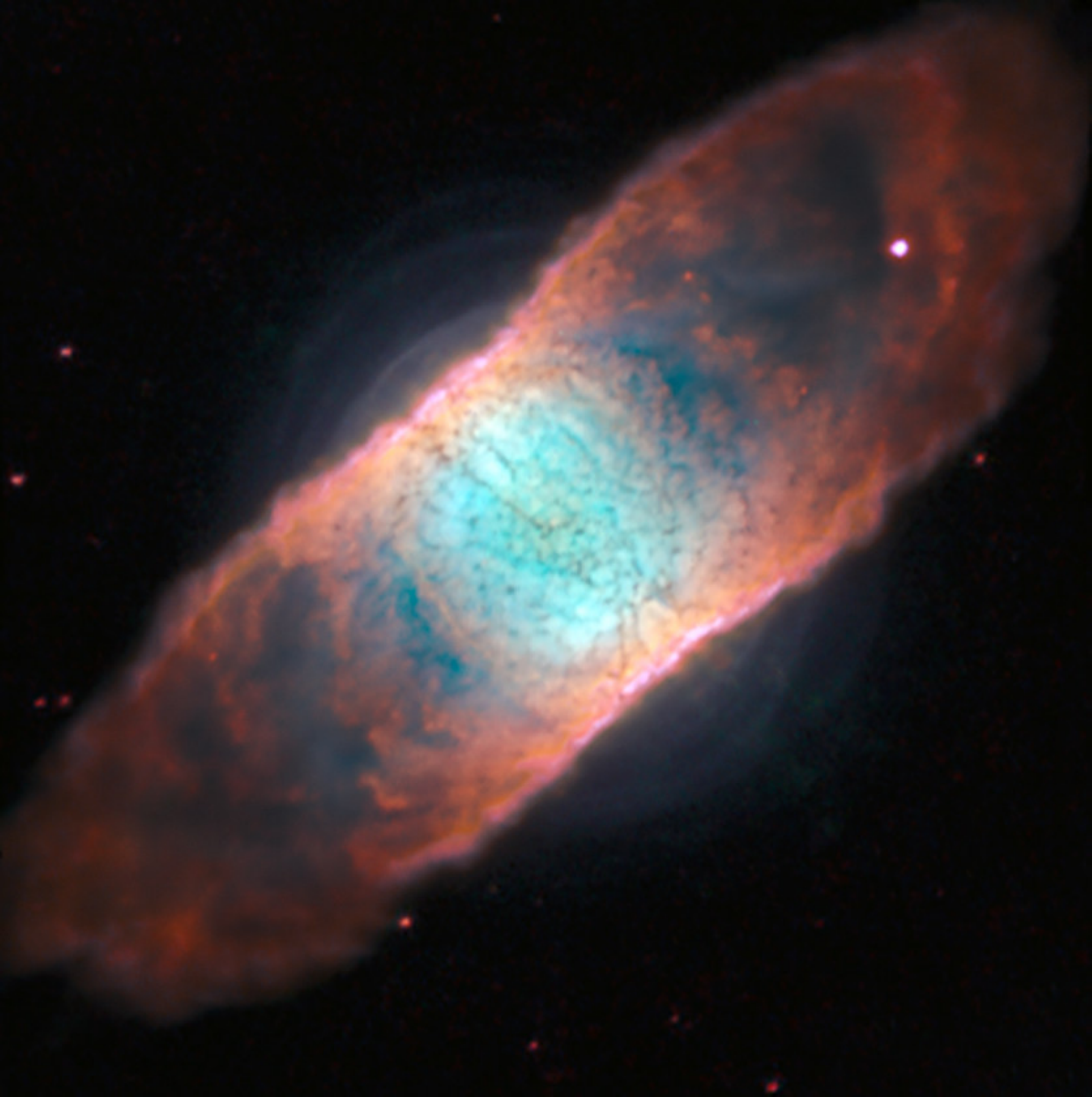}
\hspace{0.2truecm}
\includegraphics[height=5.0truecm,angle=0,clip]{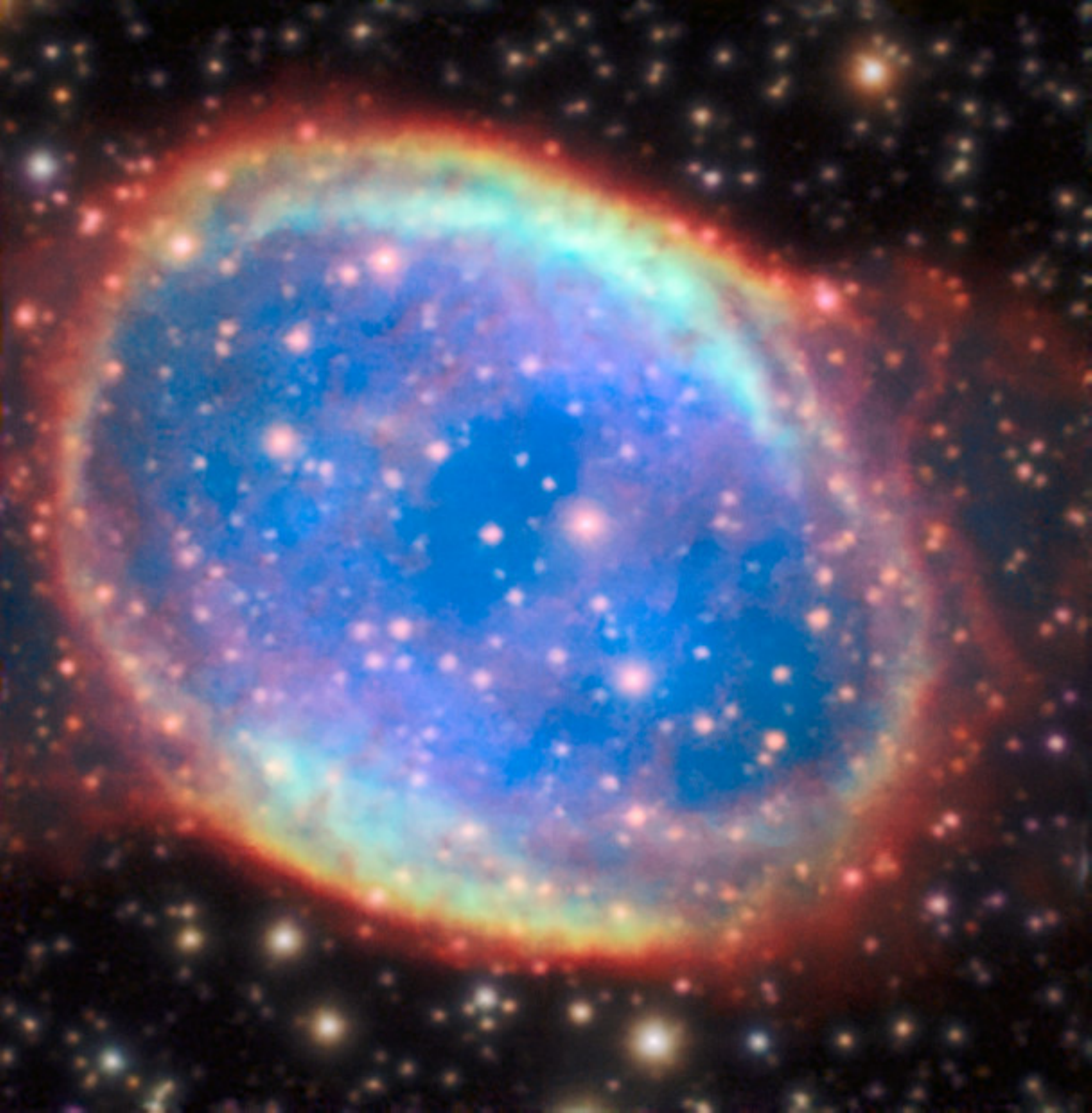}
}
\caption{Left: ESO Press Release images (eso1724) of the bipolar Galactic PN IC~4406 
(left, image size 83$\times$83$''$, north to upper left, east to lower left) and 
the elliptical PN NGC~6563 (right, image size 62$\times$62$''$, north to the top), 
both taken with MUSE as part of the commissioning of the Wide Field Mode with 
(ground layer) AO correction. The emission line images are colour 
coded according to: [O~III] blue; [O~I] magenta; H$\alpha$ green; [N~II] orange; 
[S~II] red. {\it Image credits. IC~4406: ESO/J. Richard (CRAL); NGC~6563: ESO/P. 
Weilbacher (AIP)}.
}
\label{figs:pr1724}
\end{figure}

Many PNe were observed as part of the commissioning of the instrument at
various stages (Comm. entries in Table \ref{tab:MUSE_arch}), reflecting the fact that they 
make excellent press release images and enable the capabilities of the instrument 
to be thoroughly exercised (extended and point sources, line and continuum spectra 
with many emission lines, some with fixed ratios). Figure \ref{figs:pr1724} 
shows images from ESO release (\href{https://www.eso.org/public/news/eso1724/}{eso1724})
of some of these PNe observed with
WFM and GLAO, well illustrating the quality of the imaging data that can be 
extracted from the position -- wavelength reduced cubes. IC~4406 is a Type I
PN (high He, N abundance arising from a higher mass progenitor star, 
\cite{PeimbertTorres1983}) with strong bipolar symmetry; the 
colour image, a combination of emission in the lines of [O~III], [O~I], 
H$\alpha$, [N~II] and [S~II] shows the fine scale
(dust) extinction features throughout the nebula. In contrast, NGC~6563 is an
elliptical shell PN with classical inner high ionization zone and lower
ionization shell, projected on a dense stellar field;
NGC~6563 shows a fragmented CO emission shell \cite{CoxHuggins1991}.

\section{MUSE mapping of NGC~3132 and NGC~7009}
The medium ionization PN, NGC~7009 (Saturn Nebula), was selected as
a Science Verification target for MUSE in 2014 primarily on account of its size,
which fits inside the MUSE field of view, very high emission line surface 
brightness and availability of a wealth of previous observations, including 
Hubble Space Telescope (HST) imaging \cite{Balick1998} and deep spectroscopy 
\cite{FangLiu2011, FangLiu2013}. NGC~3132 was selected from among the MUSE 
PN observations during early commissioning for detailed analysis; on account of 
its larger projected elliptical form (dimensions $\sim$ 60$\times$85$''$), a 
mosaic of MUSE pointings was performed to cover the optical emission extent. The 
ionization level of NGC~3132 is moderate with weak He~II emission, but the central 
object is a wide visual binary with an A type central-star companion to the
hot post-AGB star \cite{Ciardullo1999}.

\begin{figure}
\centering
\resizebox{\hsize}{!}{
\includegraphics[height=4.6truecm,angle=0,clip]{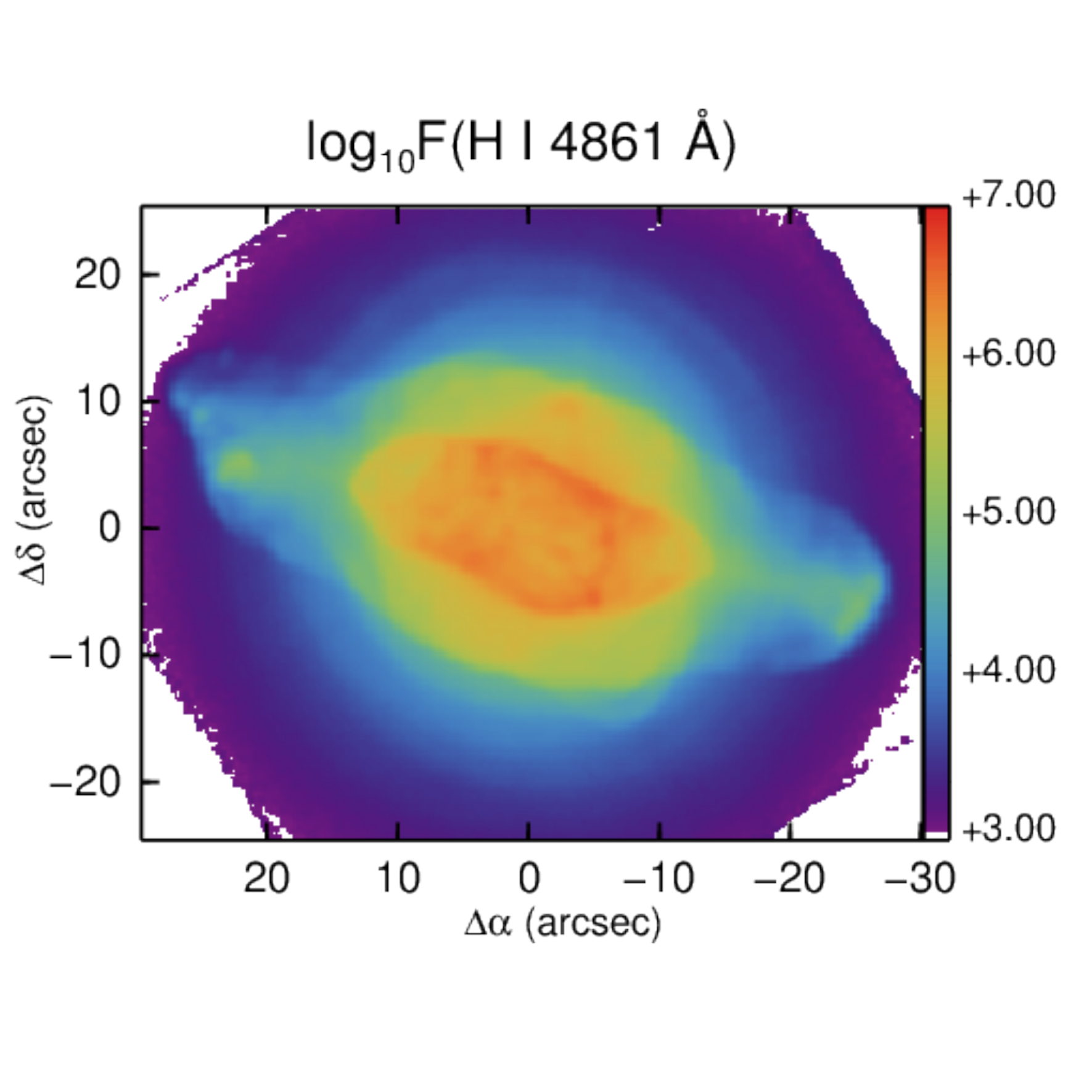}
\hspace{0.2truecm}
\includegraphics[height=4.6truecm,angle=0,clip]{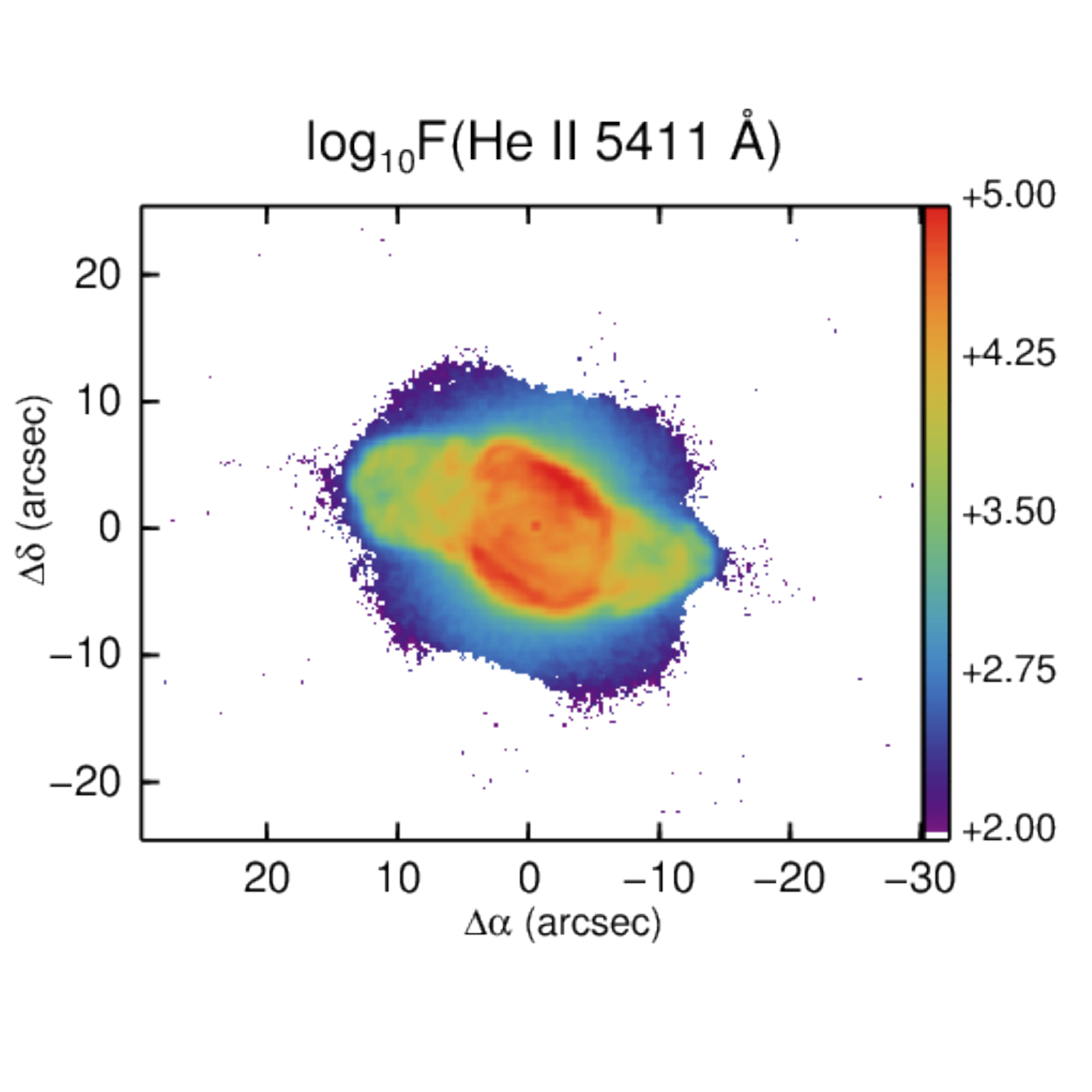}
}
\caption{MUSE emission line images of NGC 7009: hydrogen Balmer H$\beta$ 
4861\AA\ (left) and He~II 5411\AA\ 7–4 line (right), from \cite{Walsh2018}.
}
\label{figs:ngc7009}
\end{figure}

The data for both nebulae were similarly reduced and for NGC~3132 with
larger offsets to cover a full field of 124$\times$60$''$; the sky background
was estimated from the corners of the field and with a position well offset from
any nebula halo emission. Emission lines were extracted from as many spaxels as possible
given a signal-to-noise criterion and fitting multiple Gaussian profiles (see
\cite{Walsh2018, Monreal2020} for details). Short and long exposures were treated 
separately since bright lines ([O~III],4959,5007\AA, H$\alpha$, etc.) were saturated 
on longer exposures, and even [O~III]5007\AA\ was saturated over the bright shell 
emission in NGC~7009 in 10s exposure. The fitted emission line fluxes were reconstructed 
into images and Fig. \ref{figs:ngc7009} shows as an example the H$\beta$ and He~II 5411\AA\
images for NGC~7009, illustrating the difference in structure between low and high 
ionization species. \href{https://www.eso.org/public/news/eso1731/}{ESO Release 
eso1731} shows a multi-line colour composite image from these data which can be 
compared to the HST image
(\href{https://www.spacetelescope.org/images/opo9738g/}{NASA opo9738g}). 

\subsection{Spectral analysis: line ratios}

For both nebulae a common analysis was performed and the following maps were derived:
\begin{itemize}
\item reddening from ratio of H$\alpha$ to H$\beta$ (and also Paschen lines in
the red, such as P10 (9015\AA) and P9 (9229\AA)) compared to Case B 
(nebula optically thick to HI Lyman-$\alpha$ photons; \cite{Osterbrock&Ferland2006});
\item dereddened line maps (in absolute flux) for all measured lines;
\item $T_{\rm e}$ from [S~III]6312/9068\AA\ ratio, [N~II]5755/6583\AA\
and [O~I]5577/6302\AA\ for most spaxels, and from [Ar~III]5192/7135\AA\ in brighter
regions (PyNeb \cite{Luridiana2015} was used for nebular diagnostic computations);
\item $N_{\rm e}$ from [Cl~III]5517/5537\AA\ and [S~II]6716/6731\AA;
\item $T_{\rm e}$ from He~I line ratios together with 3889\AA\ optical depth;
\item $N_{\rm e}$ from the ratio among the high series Paschen lines;
\item $T_{\rm e}$ from the magnitude of the Paschen jump at 8210\AA\ with respect to 
HI P11 (8863\AA) line strength;
\item ionic abundances of many species using the appropriate $N_{\rm e}$, $T_{\rm e}$
diagnostics, such as O$^{+}$ and O$^{++}$, S$^{+}$ and S$^{++}$, Ar$^{++}$ and Ar$^{+++}$;
\item total abundances of He and O. 
\end{itemize}
For both nebulae, although these diagnostics are standard, MUSE allows for the first time
large field, high fidelity spatial mapping of their variation across the projected nebula 
surface.

\begin{figure}
\centering
\centering
\resizebox{\hsize}{!}{
\includegraphics[height=4.8truecm,angle=0,clip]{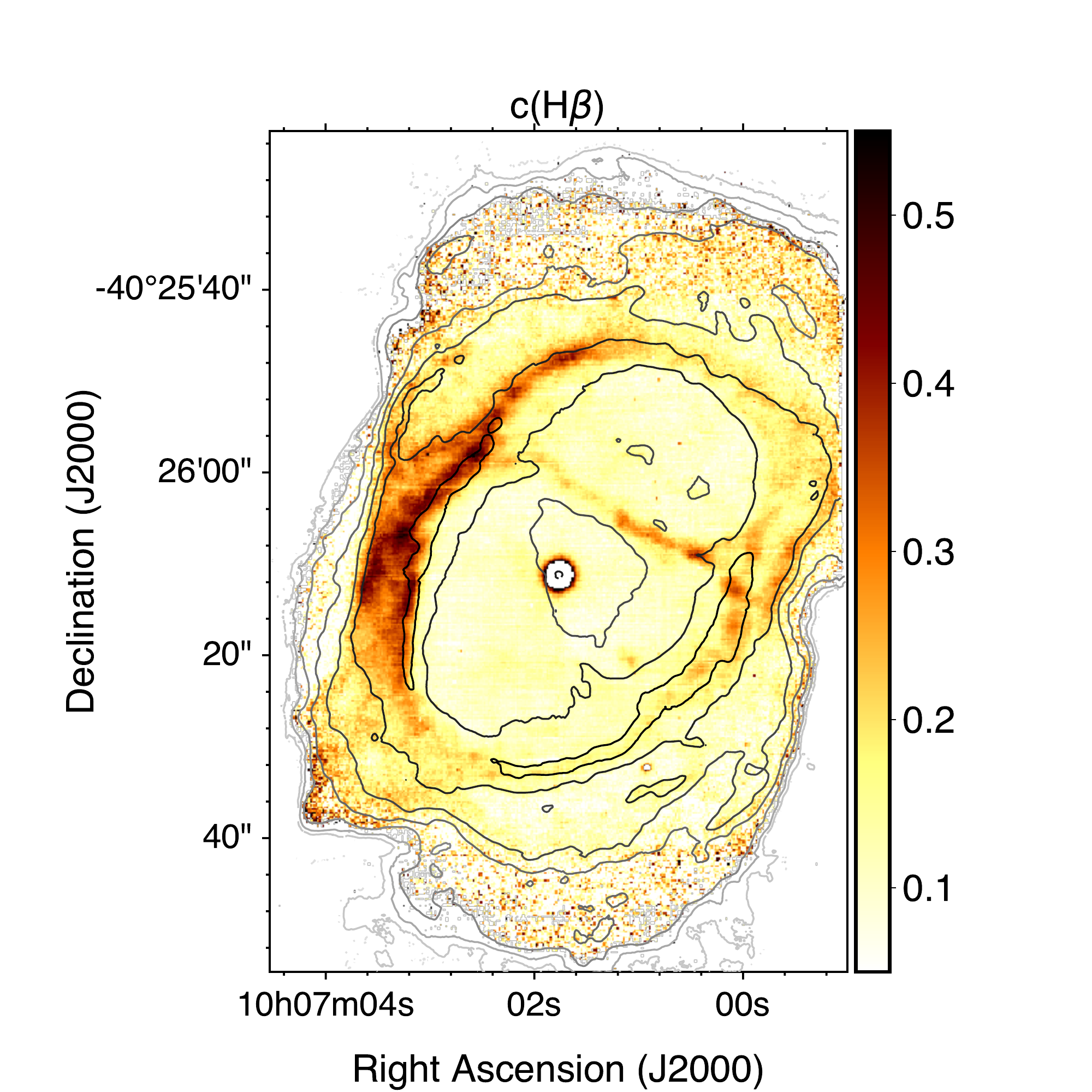}
\hspace{0.2truecm}
\includegraphics[height=4.8truecm,angle=0,clip]{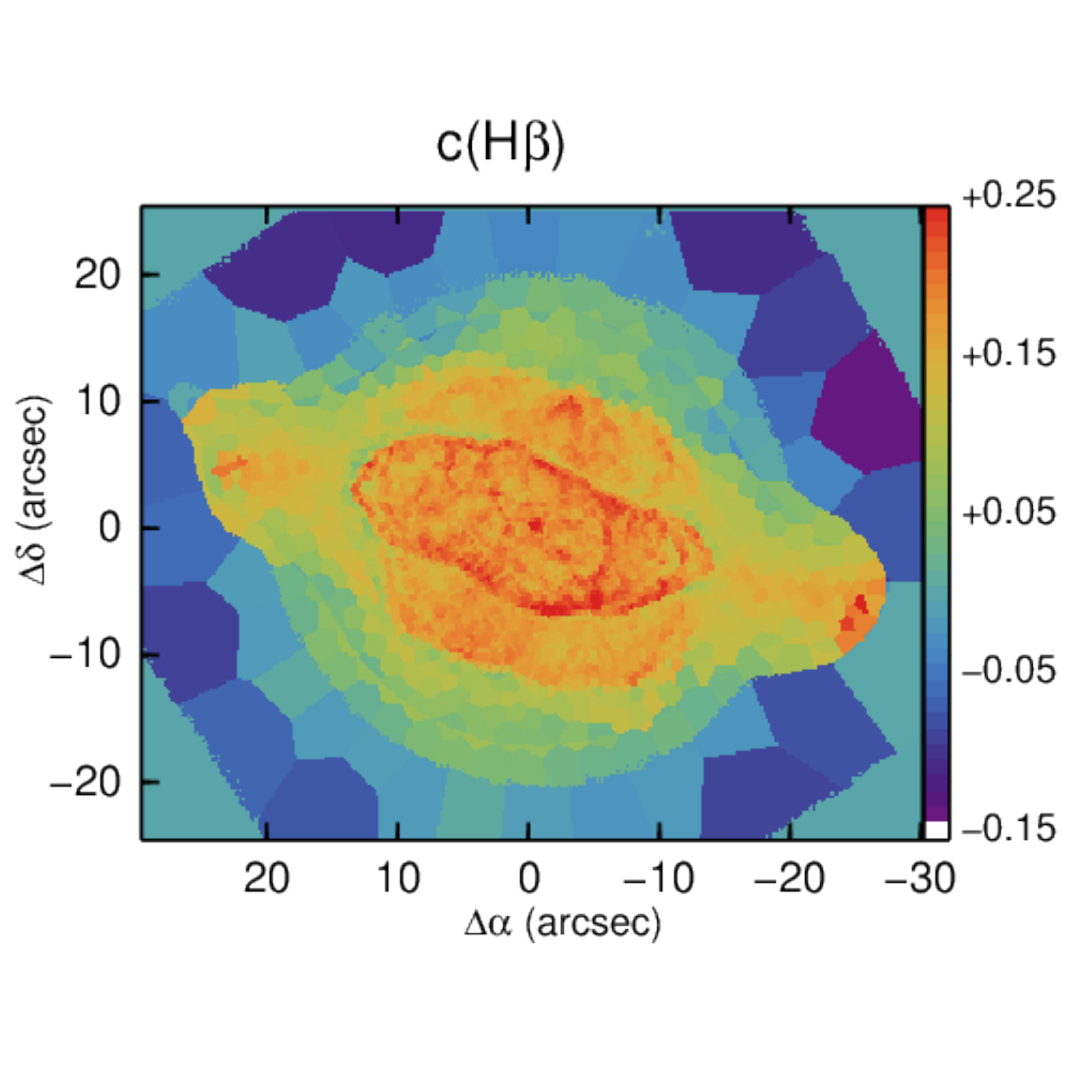}
}
\caption{Comparison of the extinction, $c$(H$\beta$), maps of NGC~3132 (left, 
from \cite{Monreal2020}) and NGC~7009 (right, from \cite{Walsh2016}) showing 
strong evidence for internal dust structures obscuring
background emission. The NGC~3132 image is produced in the native spaxel format,
whilst the $c$(H$\beta$) image for NGC~7009 is produced from the ratio of H$\alpha$ 
and H$\beta$ images which have been Voronoi tesselated to a constant signal-to-noise 
on H$\beta$ flux. 
}
\label{figs:cmaps}
\end{figure}

\begin{figure}
\centering
\centering
\resizebox{\hsize}{!}{
\includegraphics[height=4.8truecm,angle=0,clip]{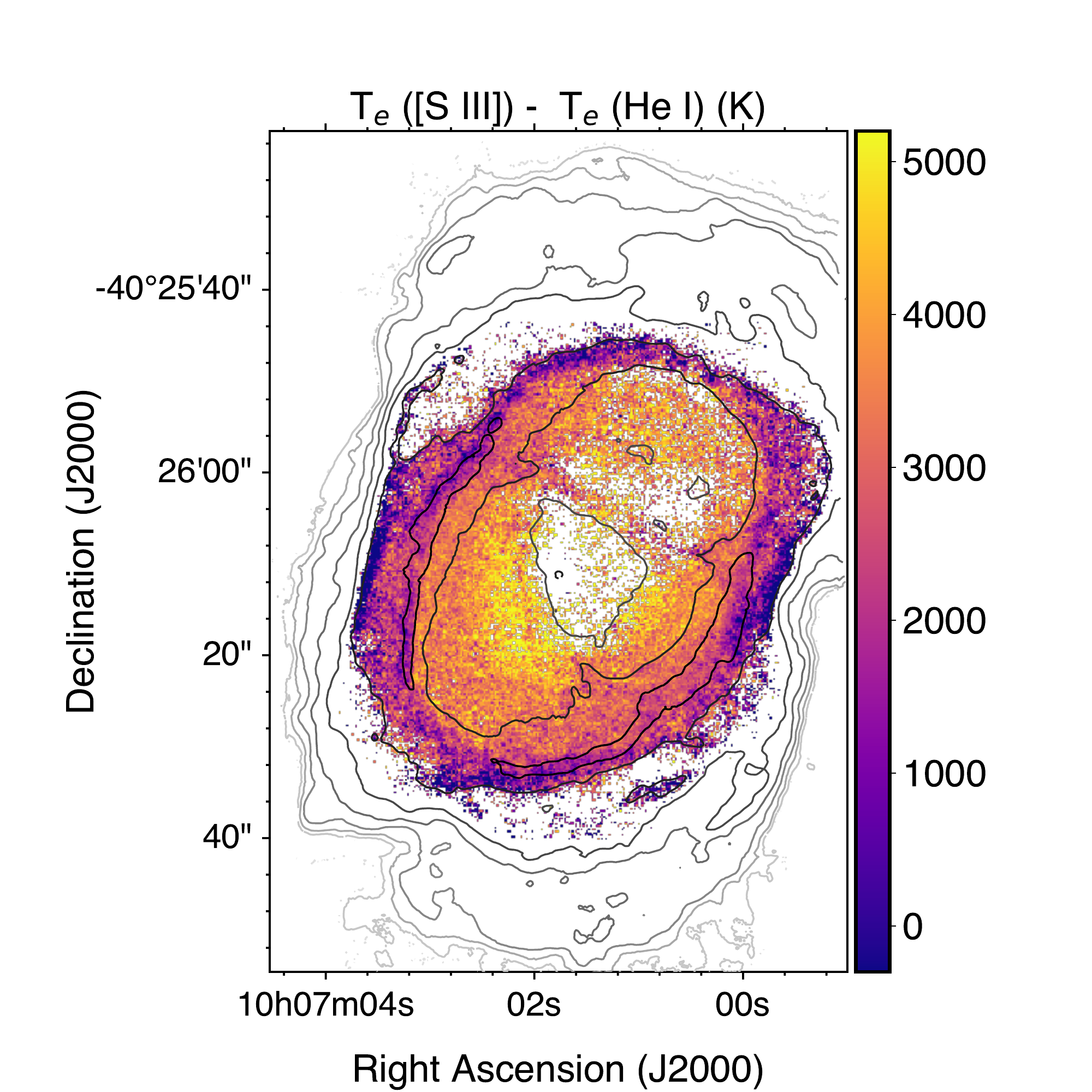}
\hspace{0.2truecm}
\includegraphics[height=4.8truecm,angle=0,clip]{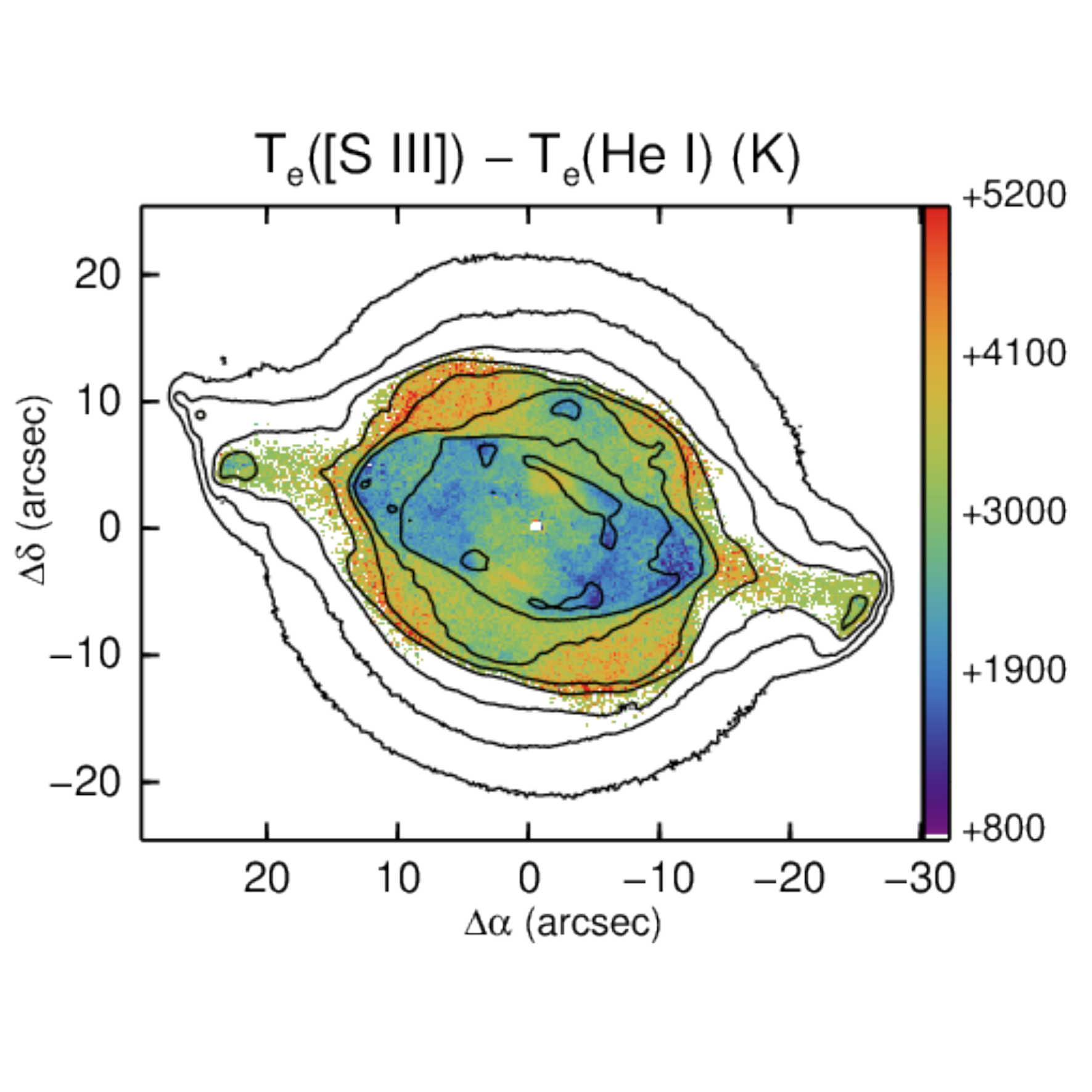}
}
\caption{Comparison of the difference in $T_{\rm e}$ between the [S~III] and He~I
$T_{\rm e}$'s in NGC~3132 (left\cite{Monreal2020}) and NGC~7009 (right \cite{Walsh2018}). 
An increasing trend (larger 
difference between $T_{\rm e}$ from [S~III] and He~I) is found in both nebulae 
towards the location of the central star within the main shell.
}
\label{figs:deltat}
\end{figure}

Whilst it is difficult to compare distinct PNe, with a sample of only two, and chosen
with diverse criteria, several points of similarity, which arise 
directly from the IFU spectroscopy, can be selected by a 
comparison of the results presented in \cite{Walsh2016, Walsh2018, Monreal2020}:
\begin{itemize}
\item the logarithmic extinction maps, $c$(H$\beta$), show distinct features over the nebulae
pointing to the presence of dust within the ionized regions (see Fig. \ref{figs:cmaps}
and \cite{Walsh2016}). Whilst not unexpected, since some evidence of spatial extinction 
has been observed in the optical for a few PNe to date, and dust emission detectable in 
the infrared is almost ubiquitous (e.g. \cite{Bernard-Salas2006}), the association of 
dust with distinct features of the nebular structure, such as shell edges, knots, etc., 
is noteworthy;
\item $T_{\rm e}$ measured from the ratio of He~I lines, or from the magnitude of the
Paschen continuum jump at 8210\AA, can be compared with the 
collisionally excited line (CEL) $T_{\rm e}$ from [S~III] line ratio, to determine
a CEL to recombination line temperature difference. In both NGC~7009 and NGC~3132,
the difference in $T_{\rm e}$ increases towards the central star (see Fig. \ref{figs:deltat}).
The radial trend of this increase is in the same sense as the increase of the 
abundance discrepancy factor (ADF) measured between ORL and CEL abundances,
for example for O$^{+}$ \cite{Wesson2018}, although not yet in the same nebulae;
\item the maps of total O abundance (i.e, O/H), derived on various assumptions of ionization 
correction factor (ICF) to correct for the fact that [O~IV] emission is not observed in
the optical, is not flat for NGC~3132 or NGC~7009 (\cite{Walsh2018, Monreal2020}, as 
would be expected from evolutionary considerations
(short term alterations in O are not predicted in the late stages of
low mass star evolution). The deviations from flatness in the O/H 
maps imply that the ICF procedure is incomplete: Figure 13 for NGC~3132 
\cite{Monreal2020} and 21 for NGC~7009 \cite{Walsh2018} demonstrate that these O abundance 
modulations are correlated with the main shells and/or ionization zones. 
\end{itemize}

\subsection{Spectral analysis: radial velocities}
Emission line profiles in PNe are broadened locally by thermal and turbulent velocities 
and globally by the nebular expansion (typically 15 -- 40 km$^{-1}$), thus line
profiles of metal lines have intrinsic widths of typically 10--20 kms$^{-1}$,
while H and He lines are broader. Although the resolution of MUSE spectra is
intermediate, around 2.7\AA\ (velocity resolution 170 -- 85 kms$^{-1}$ 
from blue to red), some limited velocity information 
can be determined from Gaussian fitting of the lines. In
\cite{Monreal2020} the velocity field of NGC~3132 was measured from several lines and shows a trend 
from higher to lower velocities from NW to SE; this trend was better seen in the forbidden
lines and for the lower ionization lines of [N~II] and [O~I], but the behaviour in the 
lower ionization species is more complex with both receding and approaching velocities in
both lobes. A simple expanding shell seems difficult to reconcile
with the pattern of velocities and the velocity field could fit much better in a 
diabolo model, as proposed in \cite{Monteiro2000}.   

\section{Prospects}
MUSE has great potential for moving spatial spectroscopic studies of PNe to a higher 
level, beyond the existing methods of slit-scanning and narrow band imaging, to
high fidelity mapping of the full extent of the nebulae from the highest
ionization regions near the central star to the neutral outer regions. There is
also great potential for discovery of additional domains in PNe, such as the 
low ionization structure (LIS) regions found in at the extremities of NGC~3132 
\cite{Monreal2020}, where a mix of photoionization and shock excitation and 
abundance imprints of distinct late-stage stellar ejecta may exist. 

Given the already substantial set of PNe observed with MUSE (Tab. \ref{tab:MUSE_arch}), 
covering a wide range in central star temperature, morphology and ADF, comparative
studies can be initiated to explore correlations between diagnostics. Such 
a database can enable improvements in the domain of different CEL ratios for 
$N_{\rm e}$, $T_{\rm e}$ measurement, with impetus for improvement of atomic data
where clear discrepancies arise. With a large set of empirical line ratios, 
diagnostics and abundances, covering a diverse set of PNe, improvements in ICF's are 
enabled with implications for all targets of nebular spectroscopy.    

The intermediate spectral resolution and lack of blue response of MUSE (Sect. 2) will
be compensated by BlueMUSE \cite{Richard2020}, with coverage to 3500\AA\ at resolving 
power of $>$4000, which is a proposed instrument for the VLT in the 2020's. 
BlueMUSE brings several strategic advantages for the study of PNe: 
availability of the auroral [O~III]4363\AA\ line for $T_{\rm e}$ measurement from the 
4363/5007\AA\ ratio for the bulk ionization stage of the dominant nebular coolant; 
coverage of the strong  [O~II]3726,3729\AA\ doublet for $N_{\rm e}$ and O$^{+}$ 
fraction; coverage and higher spectral resolution for measuring the weak O and C
ORLs, enabling the spatial mapping of the ADF, at least for O.

\acknowledgments{We thank the referees for their comments which helped to clarify the 
presentation. AMI acknowledges support from the Spanish MINECO through project 
AYA2015-68217-P.}

\authorcontributions{JRW attended the WORKPLANSII workshop in Leiden, the Netherlands 
where this work was presented. JRW and AMI each led one of the two projects described.}

\funding{This research was partially funded by Spanish MINECO through 
Project~AYA2015-68217-P.}

\conflictsofinterest{The authors declare no conflict of interest.}


\reftitle{References}

\end{document}